\documentclass[
superscriptaddress,
frontmatterverbose, 
showpacs,preprintnumbers,
nofootinbib,
nobibnotes,
amsmath,amssymb,
 aps,
twocolumn,
rmp,
]{revtex4-1}

\usepackage[english]{babel}
\usepackage[utf8x]{inputenc}
\usepackage[T1]{fontenc}

\usepackage{amsmath}
\usepackage{graphicx}
\usepackage[colorinlistoftodos]{todonotes}
\usepackage[colorlinks=true, allcolors=blue]{hyperref}

\usepackage{color}

\begin{document}

\title{PRINCIPIA: a Decentralized Peer-Review Ecosystem}

\author{Andrea Mambrini}
\email{a.mambrini@optidrome.com}
\affiliation{Optidrome Limited}

\author{Andrea Baronchelli}
\email{andrea.baronchelli.1@city.ac.uk}
\affiliation{City University of London}

\author{Michele Starnini}
\email{michele.starnini@gmail.com}
\affiliation{ISI Foundation}

\author{Daniele Marinazzo}
\email{daniele.marinazzo@ugent.be}
\affiliation{University of Ghent}

\author{Manlio De Domenico}
\email{mdedomenico@fbk.eu}
\affiliation{Fondazione Bruno Kessler}

\date{\today}                     

\begin{abstract}
Peer review is a cornerstone of modern scientific endeavor. However, there is growing consensus that several limitations of the current peer review system, from lack of incentives to reviewers to lack of transparency, risks to undermine its benefits. 
Here, we introduce the PRINCIPIA\footnote{\url{http://www.principia.network/}}\footnote{\url{principianetwork@gmail.com}} framework for peer-review of scientific outputs (e.g., papers, grant proposals or patents). The framework allows key players of the scientific ecosystem -- including existing publishing groups -- to create and manage peer-reviewed journals, by building a free market for reviews and publications. PRINCIPIA's referees are transparently rewarded according to their efforts and the quality of their reviews. PRINCIPIA also naturally allows to recognize the prestige of users and journals, with an intrinsic reputation system that does not depend on third-parties. 
PRINCIPIA re-balances the power between researchers and publishers, stimulates valuable assessments from referees, favors a fair competition between journals, and reduces the costs to access research output and to publish.
\end{abstract}

\maketitle

\tableofcontents

\section{Introduction}

Peer review is the practice of evaluating scientific output by experts in the relevant domain of knowledge. It provides a validation mechanism that is crucial to reduce the likelihood of publishing misleading results or false claims, adding value to scientific output. The impact of peer-review is pivotal at all scales of the scientific process. 
It can affect academic careers, and -- through top journals and funding agencies -- it might drive the rise or fall of academic trends or even research fields.

Despite its crucial role, the current peer review comes with many disadvantages that make it a highly debated matter across decades~\cite{Culliton1996,editorial2003reviewing,Tennant2017,editorial2019natcat}.
A remarkable example is the public criticism to funding decisions of the National Science Foundation, leading to a statistical analysis of the underlying evaluative procedures which showed no significant evidence of impact on funding~\cite{Cole1977}. 

Rationalizing peer review is challenging: the work of reviewers is essentially based on voluntary contribution with virtually no incentives to report's quality. At the same time, academic publishing is a tremendously profitable business where the workforce 1) produces valuable content for free, 2) performs valuable tasks without any reward and 3) guarantees a constant demand. 
In fact, the whole peer-review system induces a social dilemma at the individual level that is barely balanced by current incentives, such as the journal impact factor. 
Why should peers cooperate for the public good benefits? Why should they do it, if their work is done without rewards while publishers take all benefits, e.g., from subscriptions or open access policies? Why authors should be fair or produce high-quality scientific output if their peers might opt for not cooperating?

Any rational answer to such questions would lead to sub-optimal individual decisions, resulting in an unreliable scientific ecosystem and a tragedy of the commons. However, decades of results from behavioral organization and decision theory, experimental economics and survey research show that rationality is very often far from being a good model of human behavior~\cite{loomes1982regret,simon1986rationality,Tversky1989,arthur1994inductive,simon1995rationality,Kahneman2003}.

Some studies have shown that increased public good benefits, together with editorial decisions accounting for reputational information (reputation bias), can help the evolution of high-quality contributions from authors while reviewers still lack the right incentives~\cite{Righi2017}. It is emblematic the provocative study by Campanario about the unreasonable resistance encountered by some studies -- among the most-cited articles of all times -- during peer review~\cite{Campanario1996}.

Nevertheless, peer-review is still perceived by all key players, from authors to publishers, as an effective way to improve the quality of a manuscript by means of a collective effort, opposed to isolated ones by few authoritative individuals. In fact, knowledge becomes scientific only when the community reaches a consensus on it by means of academic scrutiny, refinement and validation~\cite{editorial2019natcat}.

\subsection{State of peer review}

The potential disadvantages of peer review have been widely discussed for decades. 
In 1982, 12 articles -- published by academics from prestigious and highly productive American psychology departments in highly regarded American psychology journals -- where resubmitted to the same journals which previously published them 18 to 32 months earlier, after changing authors and their affiliations. The result was that 9 out of 12 submissions were not recognized as resubmissions, while 8 out 9 those studies were rejected because of serious methodological flaws~\cite{Blissett1982}. A study similar in spirit, one decade after, highlighted the same absence of reliability in peer review of medical research~\cite{Ernst1993}.

Research interest  on peer review has recently increased, yet the field remains fragmented across different research communities~\cite{Grimaldo2018}. 
The Global State of Peer Review report for 2013--2017\footnote{\url{https://publons.com/community/gspr}}, published by a platform helping academics to track and recognize their contributions to the scientific endeavor, has highlighted that such contributions are unevenly distributed across countries and institutions. Scholars in leading science countries, such as the United States and Japan, write 2 reviews per each of their submitted articles, against the 0.6 written by their colleagues in emerging countries such as China and Poland. Moreover, the latter tend to accept more frequently peer review requests and complete them faster than the former ones, although their reviews are on average shorter than the ones from scholars in wealthy locations. According to the same report, reviewers spend globally about 68.5 million hours each year, providing short reviews (477 words on average) within 16.4 days (median value): surprisingly, 10\% of reviewers account for 50\% of peer reviews, and 75\% of editors claim the hardest part of their job is finding willing reviewers. It seems that for 41\% scholars, peer review is part of their job, although 71\% of researchers have to decline review requests because the article is not within their area of expertise and 42\% declare to be too busy. 

The excessive academic burdening, the lack of training in reviewing and the excessive number of requests, coupled to potential human biases, might lead to poor reviews which fail to prevent the publication of studies that do not deserve consideration from the academic community~\cite{editorial2019balanced}. 
One possible source of biases in the peer-review process resides in the connections -- such as co-authorship -- between creators and evaluators of scientific work. 
Interestingly, even belonging to different professional networks, or ``schools of thought”, might lead to substantive disagreements between scientists in the peer-review process~\cite{RePEc:eee:respol:v:47:y:2018:i:9:p:1825-1841}. 
One possible solution to this issue could be the adoption of  ``double-blind” peer-review, in which reviewers are not aware of the names and affiliations of paper authors, as opposed to the ``single-blind” system in which only names of reviewers are hidden. 
For papers submitted to computer science conferences, single-blind reviewers have been showed to be significantly more likely than their double-blind counterparts to recommend for acceptance papers from famous authors, top universities, and top companies~\cite{Tomkins12708}.

Even editors, as humans, tend to take decisions that might be affected by their current burdening, resulting into an overall peer review system with more disadvantages than advantages, and a scientific ecosystem supported by wrong incentives. It is emblematic the lack of policies to recognize researchers for the production of valuable data set -- even if the trend is slowly changing~\cite{Gorgolewski2013} -- and for reproducibility studies, resulting in movements against ``data lechers'' -- researchers who use data produced and shared by other colleagues -- and in defining research parasitism as a treat to science~\cite{Longo2016}. In reality, these collateral effects might be a treat only to selfish individuals, whereas the whole collectivity can only benefit from reproducibility and replicability to advance human knowledge~\cite{bergstrom2016,Greene2017,editorial2018sharing,Teytelman2018}. Recent quantitative evidence from neuroscience supports this argument~\cite{Milham2018}.

It is clear how concerns related to peer review reflect more general issues affecting the scientific ecosystem. 
Note that, despite all limitations, peer-review scores to grant applications have been showed to be at least partially predictive of the success of funded projects, quantified by their total time-adjusted citation output, while the amount of funds awarded per application is not~\cite{10.1371/journal.pone.0106474}.
Therefore, improving the peer-review process also will ultimately impact the allocation of billions of dollars in research funds.
As we will see, opening the whole peer review process, and making it transparent to anyone, might only improve the current state of the art. Remarkably, high-reputation journals and research communities are already recognizing data sharing as part of the scientific process, requiring access to data used in articles they publish~\cite{Taichman2016,McNutt2016,Gewin2016,editorial2018repro}. 

Similarly, it seems plausible to think that opening the peer-review system -- in a way that accounts for potential human biases and findings from behavioral sciences, game theory and complexity science -- might be the right way to go. Other potential benefits of opening the process include valuable examples for early-career academics and the availability of huge databases which might be analyzed by scientists to identify potential biases and propose methods to eradicate them.

\subsection{Variants and alternatives}

Since 2008 the American Physical Society recognizes its outstanding referees for their work\footnote{\url{https://journals.aps.org/OutstandingReferees}} and, more recently, Elsevier\footnote{https://www.elsevier.com/reviewers-update/story/innovation-in-publishing/elseviers-peer-review-innovations-aim-to-support-science-and-reward-reviewers} and Nature Publishing Group~\cite{editorial2019recognition} are adopting similar strategies.

Some initiatives have experimented novel and alternative schemes for peer review, from publishing reports\footnote{For instance: Biology Direct (2006), The EMBO Journal (2009), eLife (2011), F1000 Research (2012), PeerJ (2013) and Nature Communications (2016)} -- i.e., the content of the review -- and reviewers' identity~\cite{editorial2018open} -- i.e., the names of reviewers -- to crowd-based approaches~\cite{List2017}. In the latter, manuscripts are made available by editors and the crowd is given a certain amount of time to respond. Surprisingly, each paper received several comments -- even in details buried in supplementary material and supporting information -- considered informative by editors and allowing for a rapid editorial response. Quantitatively, the crowd responded much faster (days versus months) and provided more-comprehensive collective feedbacks, appreciated by authors~\cite{editorial2019natcat}.

Other initiatives like the one by the journal \emph{eLife}, similar in spirit, obtained a good response to pilot projects from the research community, while measuring higher acceptance rates for late-career researchers compared to their early- and mid-career colleagues\footnote{\url{https://elifesciences.org/inside-elife/262c4c71/peer-review-first-results-from-a-trial-at-elife}}. The European Journal of Neuroscience reported that transparent review at the end of 2016 led to better and faster reports\footnote{\url{https://www.wiley.com/network/researchers/being-a-peer-reviewer/transparent-review-at-the-european-journal-of-neuroscience-experiences-one-year-on}}. Nevertheless, researchers from different disciplines are responding differently, with more than 70\% in Evolution and Ecology opting for publishing reports when the choice is given, against the 50\% in Physics~\cite{editorial2016transparent}. A group of biologists built an online platform to keep track of this type of experiments, providing a direct way to compare trials with respect to multiple criteria~\footnote{\url{https://reimaginereview.asapbio.org/}}.

A common factor is transparency in peer-review reports, which could avoid superficial reviews or too harsh tone either from referees or from authors. 
By opening the review process, it is also possible to avoid abuses from editors and predatory behavior from publishers (e.g., the ones that do not send the manuscript for review or that accept it even with low-quality/unreliable reviews with the only goal to charge fees). 
Moreover, while many researchers are in favor of opening reviews, opening identities is seen as potentially dangerous, fearing that by disclosing names the reviewers might be incentivized to weaken criticisms or authors might be incentivized to retaliate against them~\cite{RossHellauer2017}. 
Nevertheless, the British Medical Journal reported that opening reviews and reviewers' identities, simultaneously, did not change the quality of the peer reviews, suggesting that reviewers were not intimidated~\cite{vanRooyen2010}. 
Therefore, a good trade-off might be to open reviews while asking to opt-in for open identities: anonymity, in fact, does not compromise the process~\cite{Bravo2019} and might avoid unfair use in subsequent evaluation of the authors for grants, jobs, awards or promotions.
According to a recent research, the potential benefits of published reviews are multiple~\cite{Polka2018}, from encouraging constructive comments and training examples for young researchers\footnote{https://www.elsevier.com/editors-update/story/peer-review/pilot-designed-to-help-reviewers-win-recognition-for-their-work-leads-to-better-quality-reviews,-say-editors} to preserving arguments and ideas that characterize how fields evolve, from building trust based on transparency to recognizing the important work done by reviewers.

Several journals and platforms are in the process of adopting these principles one way or another.
Pubpeer is a platform dedicated to the discussion of papers following their publication in a journal or as preprints. Even though the majority of comments address the (still very important) problem of scientific misconduct and major issues changing the conclusions of the studies, there is space for costructive discussions between authors and readers, and even public reviews of preprints. 
This aspect is crucial and several journals (including major outlets such as PLOS and ELife) are now embracing comments on papers submitted to the journals and posted as preprints, and including them in the process of editorial evaluation.
Platforms of preprint and open reviews can also serve the purpose of overlay journals, which include papers already made public, commented and revised, in  collections curated by the editorial board.

Other approaches aim at decoupling the function of journals (most importantly, dissemination) from the certification provided by peer review.
Several platforms, such as the Peerage of Science\footnote{\url{peerageofscience.org}} or RUBRIQ\footnote{\url{rubriq.com}}, offer peer-review services and then forward revised manuscript to journals, in exchange of some fee or for free. 
Other platforms, such as F1000Research\footnote{\url{https://f1000research.com/}}, combine a open peer-review system with publication services, without editorial bias. However, the major limitation of these decoupled approaches is the lack of engagement of reviewers involved in the process, that can eventually lead to the failure of such initiatives. 
As we will show in the following, we argue that principles for a open, transparent peer-review process can work better if anchored to a solid reputation system, by which participants in the system are both rewarded and responsibilized.

\subsection{Aims of PRINCIPIA}

PRINCIPIA aims at improving the current peer-review system in the following aspects:
\begin{enumerate}
\item Referees are remunerated. 
\item Remuneration of referees depends on the quality of their reviews.
\item New journals are easy and free to start. 
\item A new journal inherits the reputation of its founders (i.e. editorial board).
\item All journals are open access and cost-effective. 
\end{enumerate}

Point 1. and 2. will restore a balance of power between researchers and publishers while guaranteeing a improved quality and fairness of reviews. Point 3., 4. will increase the offer and flexibility of the journal landscape by making it extremely easy to bootstrap a new journal. Finally, point 5. guarantees a lower entry barrier both for authors aiming to publish in a journal and to scientists interested in accessing published papers.

\section{The peer-review market}

Every person using PRINCIPIA is uniquely identified by its RSA-4096 public key.

Any group of people can join and form a journal by serving as member of the editorial board. Thus, in PRINCIPIA a journal is identified by a collection of public keys. The editorial board has the following roles:
\begin{itemize}
\item It reviews the paper submitted to the journal and decide whether a paper should be accepted or not
\item It decides whether other people should be allowed to join the editorial board
\item It sets the rules of the journal (see Section \ref{sec:journal})
\end{itemize}
These actions are performed using the consensus rules explained in Section \ref{sec:consensus}. 

The PRINCIPIA framework assigns a reputation score to a journal according to the people participating to the editorial board. The mechanism is explained in section \ref{sec:reputation}. 

In order to submit a paper to a journal the authors must publish it (see Section \ref{sec:blockchain_implementation} for details on the possible implementation). Then they perform a request of publication to the journal by attaching the hash of their paper and a bid for a \textit{review fee}. The editorial board will vote using the consensus mechanism explained in Section \ref{sec:consensus-review} and decide whether to accept the bid and consider the paper for review. If the vote is positive then a set of reviewers is randomly generated from the people in the editorial board. The reviewers will perform their review and publish it. According to the rules explained in Section \ref{sec:consensus-publishing} the paper will be accepted or rejected. 
After the authors have received the written reviews they will write the final version, publish it and submit it again to the same reviewers for a final acceptance vote. The \textit{review fee} will be split between the journal's fee and an amount that will be distributed to the referees according to the mechanism awarding the good reviews explained in Section \ref{sec:consensus-review-fee}.

\subsection{Differences with the current peer-review system}

There are many substantial differences between the current peer review system and PRINCIPIA.
A journal in PRINCIPIA is uniquely identified by its editorial board. This creates a liquid system in which the reputation of a publication is identified by who was serving as editorial board in the moment of the publication. This is more flexible than the current system in which a journal's reputation is identified by its history and will help new quality journal to be created more easily since it will reduce the difficulties to bootstrap a journal's reputation. Of course established journal with a strong brand can participate to PRINCIPIA and their reputation will still be recognized outside of PRINCIPIA's system. On the other hand, inside the system their reputation will be based just on their editorial board's reputation, exactly as for any other journal.

Another remarkable difference with the current system is that referees will not work for free. They will receive part of the \textit{review fee} for the work they performed by reviewing the paper. Moreover they will be incentivized by writing detailed and high quality reviews by the system explained in Section \ref{sec:consensus-review-fee}.

In PRINCIPIA the equilibrium between authors and journals is re-established. In fact, while authors can look for the most suitable journals to publish their work, we envision a system where journals bid on papers to attract those authors of (initially perceived) high quality scientific products.

\subsection{Fee market}
\label{sec:fee_market}
In PRINCIPIA fees are not fixed and are instead decided by the market. There are two kind of fees: the \textit{review fee} which is bid by authors to have their paper reviewed by a journal, and a \textit{joining fee} which is bid by a person to join the editorial board of a journal.

Authors will have to bid a review fee in order to have their paper reviewed by a journal. This is different from the current system in which fees are paid just in case the paper gets published and the publication fee is fixed. By moving the payment at a review level we think authors will be incentivized to just submit papers on which they have enough confidence. Moreover this will discourage journals to accept as many papers as possible just to collect the publication fees. We believe this system will increase the quality of the work submitted and will free the reviewers from reviewing many low quality papers. Morover paying for reviews instead of paying for publications allows to allocate a payment to the stage where most of the work is performed (in fact the publication of a paper, after it has been reviewed, has a low cost compared to to cost of performing the review).

\begin{figure*}[!t]
\centering
\includegraphics[width=0.9\textwidth]{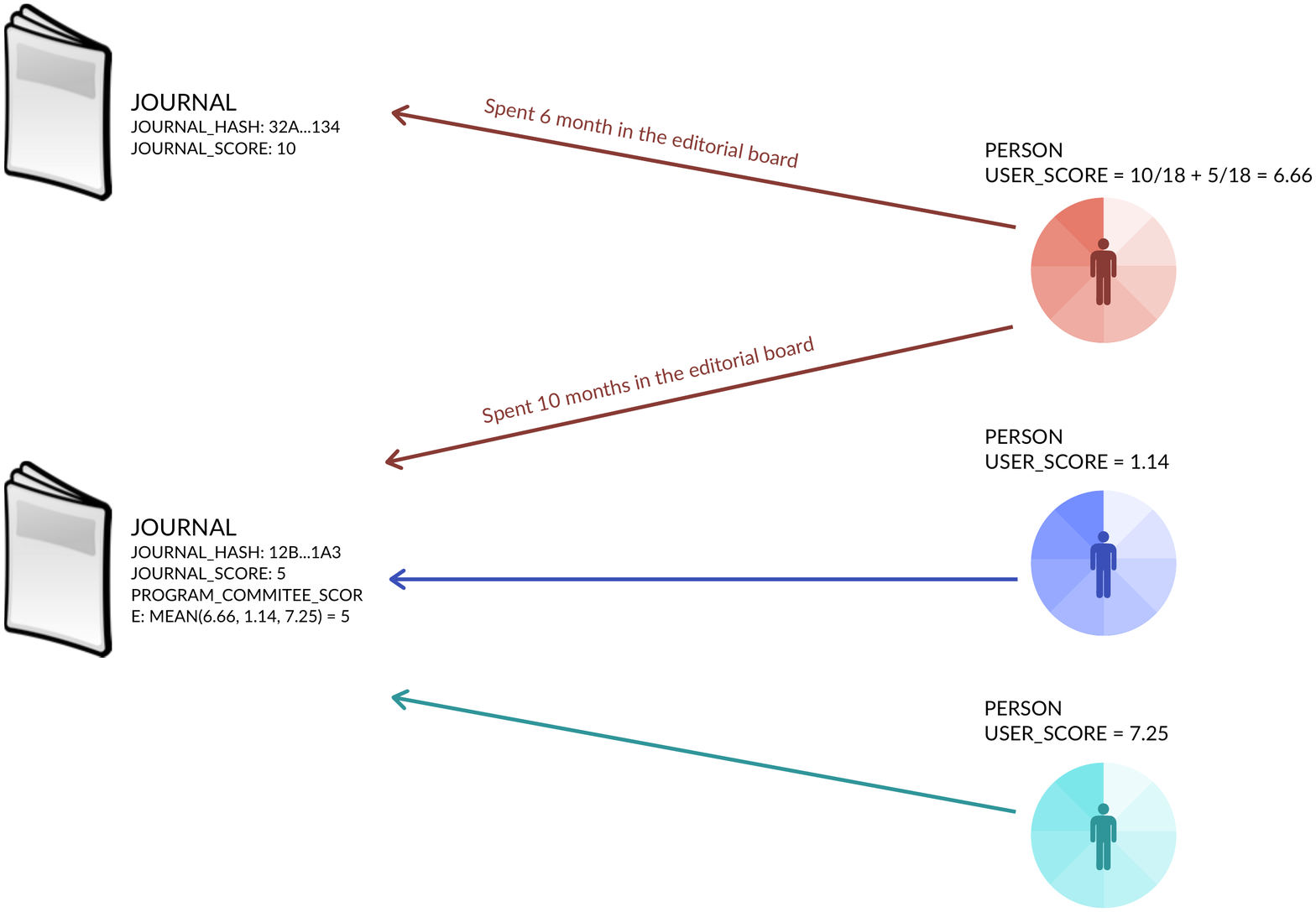}
\caption{\label{fig1} Schematic illustration of the interplay between journals and actors in PRINCIPIA. }
\end{figure*}

Keeping the review fee variable and assigned through a bid creates a market in which journal competes for the best papers. An author might propose the same quality paper to few journals and eventually decide to submit it to the journal that allows him to bid the lowest review fee (or which would even pay him to review to have the chance to review the paper). Of course this also open to the possibility for a low quality paper to be accepted for review because the authors have bid an high review fee. This is possible, but disincentivized as being accepted for review does not guarantee publication. In fact if the paper is not good, the editorial board will not be incentivized to accept it for publication as it would reduce the reputation of the journal (see Section \ref{sec:reputation}).

A similar situation happens with the fee to join a journal as editorial board member. On one hand journals migh be tempted to accept in the editorial board somebody with a bad reputation if he/she bids an high joining fee. On the other hand the reputation system will disincentivize those behaviours. Of course having a market-based joining fee will allow journal to compete for the best people to join the program committe. A person with low reputation might have to pay to join a good journal, but a person with high reputation might instead end up being paid to join the program committe of a journal.

\section{System architecture}
\label{sec:architecture}
In this section we describe in details the entity of the PRINCIPIA framework

\subsection{Person}
\label{sec:person}
A person in PRINCIPIA is a public key ($p_{pk}$). Because of the reputation system we will explain in section \ref{sec:reputation}, it is very important that each public key is strongly connected to a real person. Therefore keys need to be validated by reliable institutions, similarly to what happens in digital signature released by governments.

\subsection{Journal}
\label{sec:journal}
A journal is a collection of public keys, a reference to its \textit{ancestor journal} and the value of its parameters. The parameters of a journal are the following:
\begin{itemize}
\item Percentage of review fee to keep in the journal ($f_j$)
\item Whether reviewers should be anonymous or not ($a_j$)
\item Maximum time to perform a review ($t_j$)
\item Number of reviewers for each paper ($n_j$)
\item Size of the qualified majority needed to accept a paper for review ($r_j$)
\item Size of the qualified majority needed to spend the journal's balance. ($p_j$)
\item Size of the qualified majority needed to modify the journal ($m_j$) [this must be higher than $p_j$]
\end{itemize}

When a journal is created it requires the signature of all the members of the editorial board.

\begin{figure*}[!t]
\centering
\includegraphics[width=0.9\textwidth]{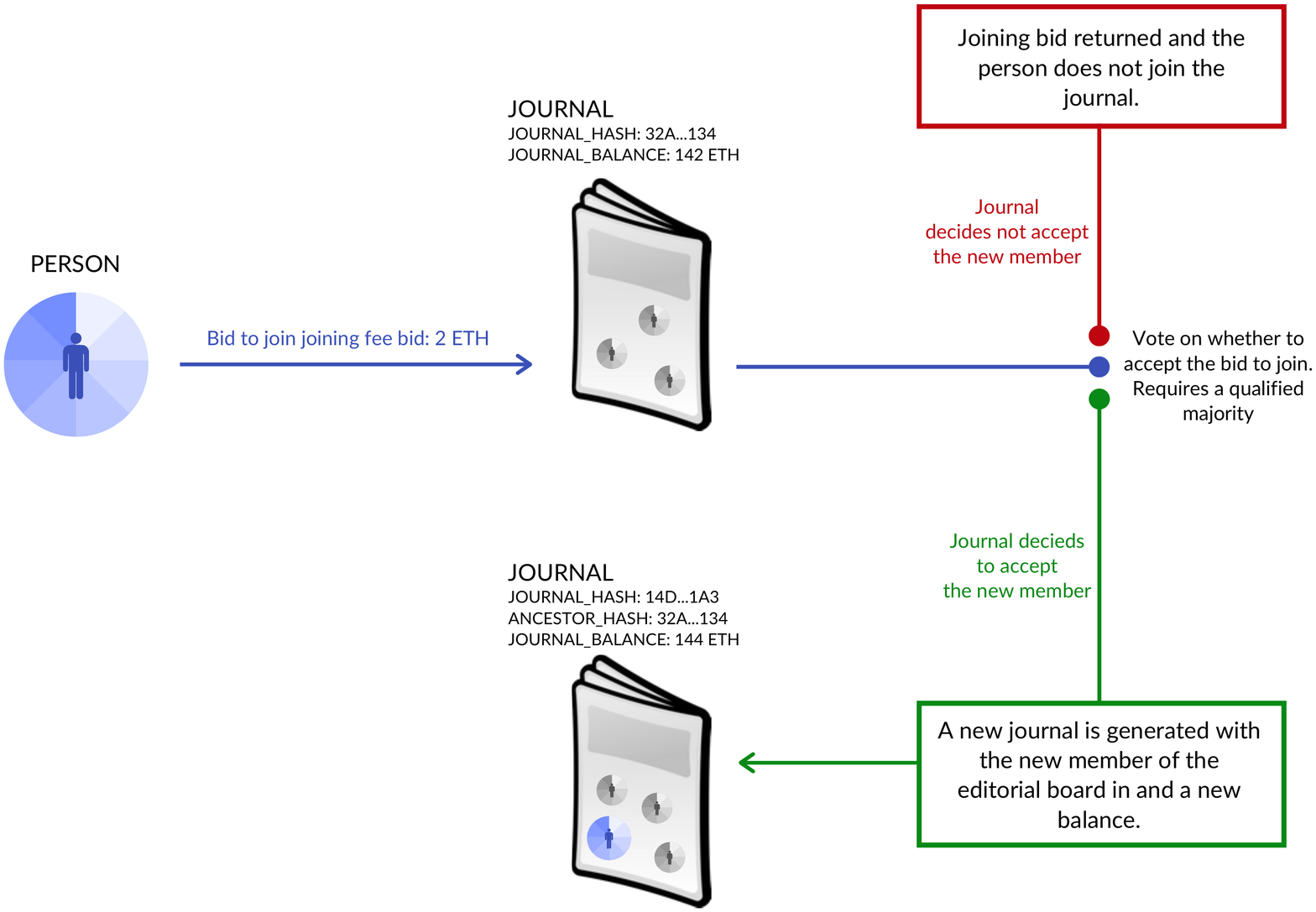}
\caption{\label{fig2} Schematic illustration of the process to extend an existing editorial board in PRINCIPIA.} 
\end{figure*}

Any of the element of the journal can be changed by creating a new journal and setting the old journal as its \textit{ancestor journal}. In order for the creation of a new journal to be valid, the signature of a certain amount of members of the ancestor journal should be present according to the rules described in Section \ref{sec:consensus}. This means that any change of the journal parameters or to the journal's editorial board triggers the creation of a new journal. It is not possible to publish on an ancestor journal, but the old publication will still be connected to the old journal. This way each publication is always connected to the status of the journal at the moment of the publication.

Each journal has a \textit{journal balance} consisting on the funds from the review fees, excluding the part which goes to the reviewers. The balance can be spent to cover the expenses of the journal (e.g advertisement or to ask a person to join the journal). Spending the funds of the journal's wallet requires the signatures of a qualified majority of the editorial board (specified by the journal's parameter $p_j$).
Since the majority needed to spend the journal balance is lower than the majority needed to modify a journal, every time a new journal is created, the balance of the journal wallet of its ancestor can (and should) be transfered to the wallet of the new journal. See Fig.~\ref{fig1} for a schematic illustration.

\subsection{Paper}
\label{sec:paper}
A papers is a published document (see Section \ref{sec:blockchain_implementation} for details on possible implementation) identified by its hash and signed by the authors.  Review and publication of a paper goes through the procedure explained in section \ref{sec:consensus}

\section{Consensus}
\label{sec:consensus}

\subsection{Modifying a journal}
\label{sec:consensus-journal-parameters}
A modification of a journal triggers the creation of a new journal which has as ancestor the old journal (as described in Section \ref{sec:journal}). Modifications are approved by a qualified majority defined as a journal parameter $m_j$.
Adding or removing a person to the editorial board is a modification to a journal. Changing one of the journal's parameters (for example the number of reviewers for each paper) is also a modification to a journal. The only journal modification which does not require a qualified majority is when a member of the editorial board decides to leave the journal, this can be done unilaterally without any approval from the journal. See Fig.~\ref{fig2} for a schematic illustration of this process.

\subsection{Joining a journal}
\label{sec:consensus-join-journal}
In order for a person to join a journal it needs to bid a \textit{joining fee}. This will trigger a journal modification event that needs to be approved by a qualified majority ($m_j$) of the journal. The amount bid will be deposited to the journal's wallet in case the person is allowed to join the journal, otherwise the amount is returned to the bidder.

\subsection{Accepting a paper for review}
\label{sec:consensus-review}
A person can ask a journal to review its paper by bidding a review fee. A paper is accepted for review by a qualified majority defined as a journal parameter $r_j$. If the paper is accepted for review, and the authors confirm he wants to proceed with the review, then $n_j$ reviewers are selected at random and they will review the paper. In case the journal want to keep the identity of the reviewers anonymous (as of the parameter $a_j$), a system based on ring signature will be used to assign the reviewers.

\begin{figure*}[!t]
\centering
\includegraphics[width=0.8\textwidth]{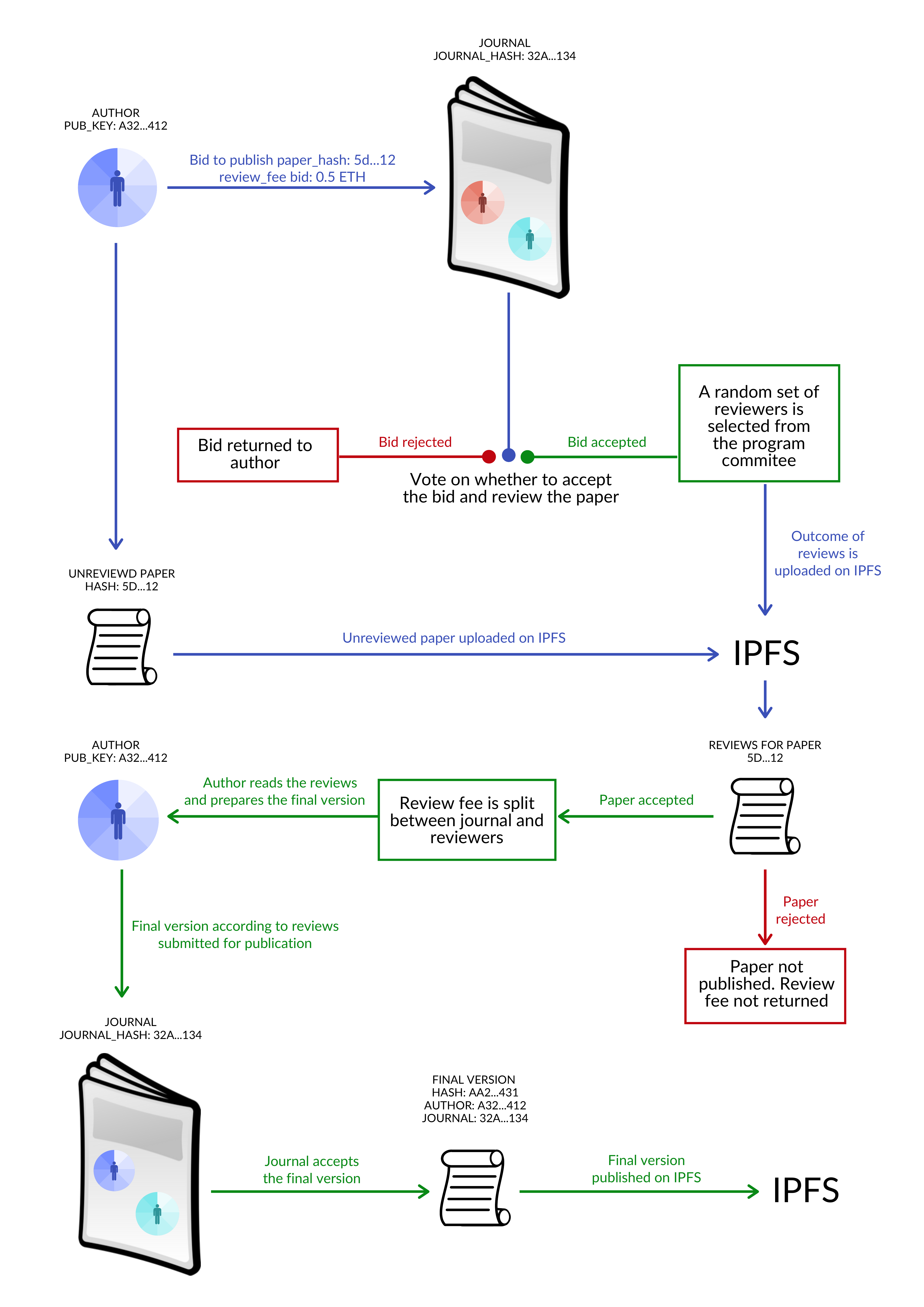}
\caption{\label{fig3} Schematic illustration of the review process leading to paper acceptance/rejection and reward for the reviewers in PRINCIPIA.} 
\end{figure*}

\subsection{Deciding whether a paper should be published}
\label{sec:consensus-publishing}
Each reviewer will review the paper and assign a score $s_u$ between $1$ and $5$. If the average score $\overline{s}$ is greater than $3$ then the paper is accepted for publication.

\subsection{How to split he review fee between reviewers}
\label{sec:consensus-review-fee}
 The review split is divided in this way:
\begin{itemize}
\item $f_j$ percentage goes to the journal's wallet
\item The remaining is split between each reviewer $u$. Each reviewers receive an amount of funds equalt to $r_u$  where $r_u$ is calculated using the following equation 
$r_u =  f_r \cdot (1-f_j) \cdot \left(\frac{1}{2}\frac{\vert s_u - 3\vert}{\sum_u \vert s_u - 3\vert} - \frac{1}{2} \frac{\vert s_u - \overline{s} \vert}{\sum_u \vert s_u - \overline{s}\vert} \right)$
\end{itemize}
The idea is to reward "consensus", i.e. to reward reviewers that agrees with the average score of the other reviewers, while at the same time discourage scores close to the $3$ and forcing reviewers to take a position. 

An illustration of the overall process is shown in Fig.~\ref{fig3}.

\section{Reputation system}
\label{sec:reputation}
Each user in PRINCIPIA has a reputation score which depends on journal in which he/she served as member of the editorial board. At the same time each journal has a reputation score which depends on the impact of the research it published. Since a modification to a journal (including change in the editorial board) triggers the creation of a new journal, which will therefore have a score that does not depends on its ancestor journal, we also define a reputation score for the editorial board, which will be the main score to judge the quality of a journal.

\subsection{Journal score}
\label{sec:journal_score}
The journal's score is inspired to the weighted impact factor described in  \cite{AlguliyevAI15}. The score is calculated as the number of citations of all the papers published in that journal divided by the number of papers published by the journal. The number of citations is weighted by the editorial board score (see Section \ref{sec:pc_score}) of journals who published the citing papers. 

Notice that a modification of the paper triggers the creation of a new journal (see Section \ref{sec:journal}) so this score takes into account just the papers published in the journal, ignoring the ancestor and descendant journals. Also notice that the weight is calculated using the editorial boarde score, rather than the journal's score.

\subsection{User score}
\label{sec:user_score}
The score of a user serving as reviewer is calculated as the weighted average of the journal's score of the journal in which he or she served as member of the editorial board. The average is weighted by the time he or she spent in the committee. So for example a user who served 6 months in a journal having journal score of 10 and 12 months in a journal having score 5, will have a score of $\frac{6}{18}\cdot 10 + \frac{12}{18}\cdot 5 = 6.66$

\subsection{Editorial board score}
\label{sec:pc_score}
The editorial board score is a score assigned to the editorial board of a journal. It is the main score to judge the quality of a journal. It is calculated as the average score of the users serving in the editorial board.

\subsection{Incentives/disincentives of the reputation system}
This reputation system gives reputation to the users according to the reputation of the journal in which they served as editorial board. This incentivize users to join journal with an high reputation.

At the same time the reputation of a journal depends on the impact of the paper published, which should encourage a journal to publish good quality papers.

The reputation of a user does not depend on the paper he/she published as an author. PRINCIPIA untight the reputation of a person as an author from their reputation as a reviewer. We want to focus on improving the peer review system and reward users who can discern high quality papers from low quality ones. We believe this is what makes a person a good editorial board member and we want the system to create a race between journals to get the best of them. We believe this will increase the quality of the work published.

\section{A minimal decentralized peer-review system}
\label{sec:alternative}

While journals are pivotal actors of the peer-review ecosystem and need to be part of it, it might be difficult to include them in the PRINCIPIA at the beginning.
Therefore, one might consider a two-step process in which i) the peer-review system is decentralized, and ii) new journals are created. In this Section, we propose an alternative  decentralized peer-review system fully compatible with PRINCIPIA, without including the formation of new journals. 

\subsection{System Architecture}
\label{sec:alt_arch}

The architecture of the system is similar to the one proposed in Section \ref{sec:architecture}. 
Each person (scientists) in PRINCIPIA is a public key. 
Each scientist in the peer-review process can have one of two roles: author of the paper submitted for review or reviewer of that paper.  

Each scientist $i$ has associated a few keywords indicating their expertise, a reputation score $RS_i$,  and a review fee $R_i$. 
The keywords of each scientist are initially validated by their publication record and updated on the basis of the papers reviewed. 
The $RS$ is initially equal for everyone and completely independent by the scientist's publication record. 
The review fee $R_i$ of scientist $i$ is the price they ask to perform a review. 
It will be dependent on their $RS_i$ (see Section \ref{sec:alt_PR}): the higher $RS_i$, the higher $R_i$ fees can be asked.

A papers is a document identified by its hash and signed by the authors. 
Authors submit a paper to the peer-review system with associated keywords indicating the field, 
 by bidding a paper review fee $R_p$.

\subsection{Peer-review process}
\label{sec:alt_PR}

Once a paper is submitted, the system tries to assign the paper to potential reviewers, by matching the $R_p$ bid by the authors with the review fees asked in the market by scientists with respect to the paper's field (i.e. also matching paper's keywords with potential reviewers' keywords).
That is, the system tries to solve the equation $\sum_{i=1}^n R_i \simeq R_p$ by choosing a pool of $n \geqslant 3$ reviewers. 
These reviewers are also chosen in order to have mixed $RS$, compatibly with the $R_p$ bid by the authors.
The system also limits the maximum number of reviews assigned to each scientists, to avoid excessive burden to single reviewers and to avoid reviewers asking low $R$ fees to get too many reviews. 
If a match is possible, the paper is accepted for review and assigned to the reviewers. 

The peer-review process proceeds as it follows:
In step 1, selected reviewers perform review independently and assign the paper a score, representing its quality. 
In step 2, selected reviewers read the reports of other reviewers and assign them a score, judging the quality of the reports. 
For each reviewer $i$, its average report score is computed. If it is above a certain threshold, the reviewer collects the $R_i$ fee asked. If not, that $R_i$ fee is returned to the paper’s authors. 
In this way, there are no incentives to give low or high scores to  reports of other reviewers to collect their fees.
Then, the reputation score $RS$ of each reviewer is updated with the average score obtained by their report. In this way, reviewers have incentives to produce high-quality reports, see Section \ref{sec:alt_inc}.  
Note that step 2 is a further burden for reviewers with respect to the current peer-review process, but relatively light since reviewers have already reviewed the paper in step 1 and so they can easily judge other report's quality.  

Once step 2 is concluded, the average peer-review score of the paper $S_p$ is computed. 
If it is above a certain threshold, which can be different for different fields, the paper is accepted by the system as ``peer-reviewed”. Otherwise it is withdraw and disappears. 
Further interactions among the authors and the reviewers could be considered, as in the current peer-review process, to achieve consensus, if necessary. 

Once a paper is accepted by the system, it becomes public together with its associated score $S_p$, the (anonymous) reports, the scores assigned to each report indicating its quality, and the reputation score $RS$ of the $n$ reviewers. 
These elements (paper's score, reports, reports' scores, and reputation scores of reviewers) certify the quality of the paper. 
At this point, authors can submit the paper for publication to journals and journals can bid to publish papers they found interesting, based on the quality of the paper certified by the peer-review system.
The authors can also decide to leave the paper in the system and not to publish it in any journal.

\subsection{Reputation system and incentives}
\label{sec:alt_inc}

The $RS$ of each scientist is at the core of the peer-review system. It represents their prestige as a reviewer, in the same way the h-index or number of citation might represent their prestige as an author. Note that these two quantities are completely independent. 

Scientists have the incentive to increase their $RS$ in order to ask higher $R$ fees, see below.  
 As described in Section \ref{sec:alt_PR}, the $RS$ increases for each review performed by the average score collected by the report of the reviewer. Therefore, reviewers have two incentives to produce high-quality reports: i) to collect the $R$ fee (which is not payed for low-quality reports), and ii) to obtain high scores and increase their $RS$. 
 These incentives are expected to solve one of the most important problems of the current peer-review system, low-quality reports. 
  
Coherently with PRINCIPIA, the $R$ fees and $R_p$ bids are not fixed but decided by the market, that is, reviewers (authors) can ask (bid) whatever fee they like.
The peer-review system is an actual market that matches the demand from peer-review service from authors to the supply of peer-review service provided by reviewers.

Reviewers have the incentive to perform reviews to increase their $RS$. In order to obtain papers for review, they have the incentive to ask low $R$ fees to be easily matched with the $R_p$ fees offered by a paper's authors. 
Note that the matching is crucially dependent on the $RS$ of reviewers: if the $R$ fee asked by a reviewer is not aligned with their $RS$, they do not obtain any paper to review. 
Therefore, reviewers have the incentive to ask a $R$ fee aligned with their  $RS$. 
Once their $RS$ increases, they can ask for higher  $R$ fees. 
Note that the system limits the maximum number of reviews for scientist, so none has incentives to ask too low  $R$ fees. 
To sum up, reviewers will ask a $R$ fee determined by the market for their $RS$ and field of expertise (associated keywords). A $R$ fee asked higher than the market value would not get any paper to review. 

At the same time, authors have the incentive to bid high $R_p$ fee to attract reviewers with high $RS$, who can charge high $R$ fees. However, note that this does not guarantee a positive peer-review outcome, with a good paper's score. 
Again, authors will bid a $R_p$ fee determined by the market, depending on the quality of the reviewers they want, certified by the $RS$ of the reviewers. A $R_p$ fee bid lower than the market value would not obtain any reviewer. 

To help the matching process, the system can suggest a ``fair" $R_p$ bid to associate to the paper in order to secure reviewers, based on the current $R$ fees asked in the market for the paper's specific field.

\section{Implementation as a blockchain-powered solution}
\label{sec:blockchain_implementation}

Implementating PRINCIPIA requires a solution to the following issues:
\begin{itemize}
    \item How to process the payments for fees (review fee and journal joining fee)
    \item Where to store papers and reviews
    \item How to implement and where to run the implementation of the protocol 
\end{itemize}

A possible solution would be to set up a no-profit foundation which would handle all the aspects above. The foundation would develop the code, host the platform, deal with the storage of papers and review and process all the payments in the main currencies. This would be similar to the model behind Wikipedia.

Such a centralized approach will expose non-trivial financial problems as the foundation would have to process large amount of money in fees. Morover having the whole academic publishing system in the hand of a foundation might lead to conflicts of interest in the way the foundation is managed which might lead to censorship, sabotage and lack of transparency. 

We think that PRINCIPIA would benefit from an implementation running on platform enabling smart contracts, like for example Ethereum~\footnote{\url{https://github.com/ethereum/wiki/wiki/White-Paper}} and from using a decentralized storage to store papers and reviews, for example IPFS~\cite{Benet2014}, a peer to peer distributed storage system in which files are identified and retrieved by hash

If the protocol is implemented using smart contracts on Ethereum, all the payments can be automatically processed using any token supported by the Ethereum platform (including Ether) so the foundation would not have to deal with processing the payments. 
Moreover the process would be completely transparent as the code of the smart contracts could be inspected and all the steps of the protocol would be public and indefinitely stored on the blockchain.

Using IPFS for storage would allow to cut down the cost of storage and lead to a system which is more censorship resistant compared to a centralized approach.

With a blockchain-based solution a foundation would still be needed in order to fund the development of the system and support institutions in starting to adopt the use of PRINCIPIA. On the other hand on the long term the foundation would have no control on the system therefore avoiding centralization of power and conflict of interests.

Blockchain solutions also come with drawbacks, e.g. performance, cost of development and flexibility \footnote{\url{https://arxiv.org/abs/1904.13093}}. Blockchain should be used just when specific trust issues need to be addressed. We envision PRINCIPIA becoming the system on which the whole peer review system can take place, therefore we think the additional complexity which a blockchain-solution would bring is balanced off by the system becoming censorship resistant, permissionless, transparent and decentralized.  

According to a recent report~\cite{Science2017}, some of the most prominent issues currently faced by scholarly communication -- e.g., costs, openness, and universal accessibility to scientific information -- might be solved by adequate processes based on blockchain. In fact a number of different blockchain-based for peer reviews have been proposed recently. 

Pluto networks~\footnote{\url{https://pluto.network/}} aims at providing an ecosystem in which research can be published and reviewed. Papers can be published by any actor in the system, and journals do not exist. We think this process has limitations as it opens up to the possibility for a reviewer to review a work outside of her expertise. Moreover removing completely the concept of journal would not allow the current editors to participate in the system, leading to difficulties in bootstrapping the system.

ScienceRoot~\footnote{\url{www.scienceroot.com}} aims at defining a system for research funding and publications based on blockchain. While having some elements in common with PRINCIPIA, ScienceRoot does not tackle some of the problems which PRINCIPIA aims at solving. There isn't a reputation system so the problem of concentration of powers in few journal is not solved. Moreover ScienceRoot lack a system of incentives to incentives good reviews and the creation of good journals.

Blockchain For Peer Review~\footnote{\url{https://www.blockchainpeerreview.org/2019/03/a-deeper-dive-into-our-proof-of-concept/}} is also developing a blockchain-based system for peer review. Also this system does not tackle all the problems PRINCIPIA is aiming to solve, for example there isn't a liquid reputation system, and the system is also lacking a process for renumerating reviewers.

Ledger Journal~\footnote{\url{https://ledgerjournal.org/ojs/index.php/ledger}} is a peer-reviewed journal which publish research articles on the subjects of cryptocurrency. Ledger is not aiming at replacing the whole peer review system, despite that it contains some interesting elements which are including in PRINCIPIA as well. For example papers are digitally signed by the author and publication timestamp is published on blockchain. Moreover reviews are published together with the paper.

PROBO~\cite{probo2017} is a blockchain-based solution to the issue of reproducible data analysis. Similarly to ScienceRoot and PLUTO, PROBO does not provide for a system of incentives/disincentives to award good behaviour.

Some criticism on using blockchain to improve the scientific process are discussed in~\cite{Extance2017}. The work is mainly critical on using blockchain to store and collect data as ``Costs in research applications would increase faster than it does for cryptocurrencies because modern science produces far more data.``. 
PRINCIPIA does not aim at publishing any data on the blockchain as it focus only on providing a system for transparent peer review. Storage on blockchain is extremely expensive and PRINCIPIA will store on the blockchain only the outcome of reviews and all the transactions. The reviews itself and the papers will be stored on a distributed storage system therefore the cost of maintaining the system can be limited.  
In PRINCIPIA the cost for publishing increase only if the number of submission increases more than the number of reviewers. Since part of the review fee will go to reviewers this will incentivize more reviewers to join the system when the cost of reviewing increases, therefore re-balancing the demand and offer in a way similar to the dynamic pricing system used in the sharing economy applications (Uber, Airbnb, Deliveroo, ...)
The article also suggest the use of a private permissioned blockchain to reduce costs and power consumption. This is a problem that does not concern specific applications but it is rather a tradeoff between security and power consumption which affect most of the current blockchain implementations and it is out of the scope of this paper. Private blockchains do not solve entirely the problem of trust and therefore are not a viable solution for the implementation of a system like PRINCIPIA.
The article also mention the problem of immutability of storage as what is published on the distributed storage cannot be deleted. This can be problematic when sensible data is published, which is not the case of PRINCIPIA as only reviews outcomes and transactions will be stored on the distributed storage.

\section{Conclusion and outlook}

The public debate around the peer-review system thrives, spurring a plethora of new proposals to improve the current model and overcome its limitations. 
Interestingly, some of these proposals are implemented by blockchain technology \footnote{\url{https://github.com/aletheia-foundation/aletheia-whitepaper/blob/master/WHITE-PAPER.md\#a-blockchain-journal}}, as we suggest here. 
Far from covering the extensive, fast-growing literature in the field, this white paper presents a novel framework for peer-review, open to contribution from all key players of the scientific ecosystem.
Our proposal is built around few principles aimed at creating a market for peer-review services and publications. 
PRINCIPIA allows reviewers to be rewarded, thus improving the quality of their reviews, and opens the publishing market by lowering entry  barriers. 
While we solidify these principles in a coherent architecture, the details of such implementation remain open for discussion.

We propose two different yet partially compatible implementation of PRINCIPIA, a whole system including a detailed description of how new, liquid journals work, and a minimal implementation of the peer-review system independent by the creation of new journals. 
Both proposals are built around two key ingredients: an intrinsic reputation system and a decentralized market for peer-review services. 
These ingredients are aimed at overcoming the main limitations of the current system, as well as potential limitations of other frameworks.  
As discussed in the introduction, indeed, several proposals include -- at least partially -- the principles presented here, in particular the idea of a transparent, open peer-review system. 
While some are thriving, others, such as Axios Review, failed due to a lack of engagement of scientists involved in the peer-review process.
We believe that the intrinsic reputation system proposed here, which reward and responsibilize reviewers, can engage scientists in joining PRINCIPIA. 
In the long term, the reputation gained through reviews could form part of the overall reputation of a scientists, in the same way citations are related to reputation as an author. 

The creation of a market for peer-review, on the other hand, will incentivize i) reviewers to improve the quality of their reports, being rewarded for it, and ii) authors to judge the quality of their papers by bidding an appropriate review fees. 
This is another interesting difference with respect to the current system, where the rational for authors is always to submit a paper to journals with the highest reputation, since fees are asked only in case of acceptance, while peer-review process is not paid -- at least in the majority of STEM-related journals. 

The two proposed architectures implement these ingredients in slightly different ways, each one with its own strengths and limitations. 
The whole system including journals anchors the reputation of users to the reputation of journals they serve, which in turns depends on the citations collected by papers published on such journals. Therefore, the reputation of a user is ultimately dependant on the quality of papers published in the journal they serve. 
This mechanism is similar to the current system in principle, while the differences root in the fact that journals are now liquid and much more easy to start. 
Therefore, it is expected that the journal market, where users are free to join and leave journals at any time, stimulates users to join journals offering better peer-review service, which will thus attract better paper, ultimately increasing the reputation of users joining them.  

In the minimal version of PRINCIPIA independent by the creation of new journals, on the contrary, the reputation system is endogenous to the peer-review market and thus completely independent by the quality of papers published, depending only on the quality of the reviews.  On the one hand, this endogenous mechanism is at risk of being self-referential, based on the fact that reviewers assign fair scores to the reports of their colleagues. 
This is however the same principle of peer-review itself, reviewers are trusted to fairly evaluate papers. 
On the other hand, the fact that the reputation score of users is directly related to the quality of their reports incentives the production of high-quality reports. 

The functioning of the peer-review market is also different between the two proposals. 
While in both cases authors are encouraged to bid a review fee depending on the quality of their paper, the mechanisms ensuring this incentive are slightly different. 
In the whole system including journals, authors submit their papers to journals, thus the review fee offered must be aligned to the reputation of the journal. 
A possible limitation here is the incentive for journals to accept many submission to collect the review fee (since this is not refunded in case of rejection), and then publish only good papers, to keep a high reputation. 

In the minimal peer-review market, on the contrary, authors submit papers directly to the market, thus the quality of the reviewers secured (determined by their reputation score) directly depends on the review fee offered. 

The proposal for a minimal peer-review system stems from the fact that the whole system including journals requires first scientists to create journals, and second the peer-review market to start, since reviewers are chosen among the editorial board of journals. 
Therefore, the creation of a journal market precedes -- at least conceptually -- the creation of the peer-review system. 
This is a possible limitation, since in order to join journals users need to bid a journal fee. 
This can be seen as an investment: the higher the journal fee offered, the better the journal joined and thus higher review fees are collected in the future as a member of the editorial board (when acting as a reviewer). 
Therefore, the process of creating new, liquid journals through the journal market could be slow, also because being part of the editorial board of a journal requires additional work with respect to peer-review (accepting/declining papers and potential new members).

A possible solution that combines the two proposal is to start first the minimal version of the peer-review system, then once reputation scores of scientists start to increase and diversify, liquid journals can be created by scientists bidding their reputation score. 
In this way, good reviewers would join and form high-quality journals.

\bibliographystyle{apsrev}
\bibliography{Bibliography}

\end{document}